# DeepOrientation: convolutional neural network for fringe pattern orientation map estimation

Maria Cywińska[1,2], Mikołaj Rogalski[1], Filip Brzeski[1], Krzysztof Patorski[1], Maciej Trusiak[1,3]

[1]*Warsaw University of Technology, Institute of Micromechanics and Photonics, Faculty of Mechatronics, A Boboli 8, 02-525 Poland*
[2]*maria.cywinska.dokt@pw.edu.pl*
[3]*maciej.trusiak@pw.edu.pl*

**Abstract:** Fringe pattern based measurement techniques are the state-of-the-art in full-field optical metrology. They are crucial both in macroscale, e.g., fringe projection profilometry, and microscale, e.g., label-free quantitative phase microscopy. Accurate estimation of the local fringe orientation map can significantly facilitate the measurement process on various ways, e.g., fringe filtering (denoising), fringe pattern boundary padding, fringe skeletoning (contouring/following/tracking), local fringe spatial frequency (fringe period) estimation and fringe pattern phase demodulation. Considering all of that the accurate, robust and preferably automatic estimation of local fringe orientation map is of high importance. In this paper we propose novel numerical solution for local fringe orientation map estimation based on convolutional neural network and deep learning called DeepOrientation. Numerical simulations and experimental results corroborate the effectiveness of the proposed DeepOrientation comparing it with the representative of the classical approach to orientation estimation called combined plane fitting/gradient method. The example proving the effectiveness of DeepOrientation in fringe pattern analysis, which we present in this paper is the application of DeepOrientation for guiding the phase demodulation process in Hilbert spiral transform. In particular, living HeLa cells quantitative phase imaging outcomes verify the method as an important asset in label-free microscopy.

## 1. Introduction

The full-field optical measurement techniques, such as interferometry [1-3], holographic microscopy [4-6], fringe projection [7,8] or moiré technique [9], are considered to be highly accurate, non-invasive and fast ones. In all mentioned techniques the measurement result is received in the form of a fringe pattern (interferogram/hologram/moirégram), where the phase function (or less frequently amplitude function) stores information about studied specimen. For that reason, the whole process resulting in information retrieval from recorded fringe pattern can be divided into two steps: opto-electronic measurement leading to capturing the fringe pattern and numerical processing leading to the fringe pattern phase map calculation. In general, recorded fringe pattern can be described as:

$$I(x,y) = a(x,y) + b(x,y)cos(\varphi(x,y)) + n(x,y), \qquad (1)$$

where $a(x,y)$ describes background intensity, $n(x,y)$ represents noise, $b(x,y)$ and $\varphi(x,y)$ denote amplitude and phase modulation (measurand), respectively.

There are generally two main classes of algorithms enabling phase map demodulation, i.e., multi- and single-frame methods. The first one is known as the most accurate, but difficult to apply in the case of studying transient events or performing measurement in an unstable environment, as generally large number of frames is needed (3+). Because of that the development of single-frame algorithms is needed and important. The Fourier transform (FT) method [10] is a well-known representative of such a technique but it has limitations in terms of the carrier spatial frequency and global spectrum filtering. The FT localized relatives, such as the windowed Fourier transform (WFT) [11], continuous wavelet transform (CWT) [12] and empirical wavelet transform [13], or other approaches including spatial carrier phase-shifting (SCPS) [14], and regularized phase tracking [15], are generally very capable but require a set of parameters to be fixed. They can be computationally and algorithmically demanding, and exhibit characteristic errors (e.g., the CWT method introduces errors in areas of strong phase gradients correctable for an especially tailored numerical scheme). Other solutions escaping so-called off-axis interferogram regime are Kramers-Kronig relation [16], Riesz transform approach [17,18], Hilbert Phase Microscopy [19-21] or two-frame Hilbert transform approach [22]. The approaches based on Hilbert spiral transform (HST) [23-25] enable the single-frame phase analysis in the widest range of fringe pattern carrier frequencies, however they do need the fringe orientation map for guiding the phase demodulation process.

It is to be highlighted that the fringe orientation map is essential in various fringe processing and analysis tasks, where it enables or greatly enhances the calculations. The examples are: fringe filtering (denoising) [26-43], fringe pattern boundary padding [41,44], fringe skeletoning (contouring/following/tracking) [27,29,32,33,36,37,39,40,44–46], local fringe spatial frequency (fringe period) estimation [30,34,47,48] and fringe pattern phase demodulation [28,30,32,36,38,47–59].

To be precise we would like to introduce the concept of local fringe direction (LFD) map and explain the difference between local direction and orientation maps. The LFD map ($\beta(x,y)$) stores the information about the azimuth of vector locally normal to fringes as well as its direction (e.g., up or down for vertical azimuth). It is a modulo $2\pi$ indicator, therefore. The LFD map cannot be calculated in the straightforward way from recorded pattern as carrier fringes with opposite directions visually are the same. The quantity, which we can calculate directly from the fringe pattern is called fringes orientation (FO) [60] and it is a modulo $\pi$ indicator. It stores the information only about the azimuth of the vector locally normal to fringes. To move from the fringes orientation to fringes direction one needs to apply the unwrapping procedure (with the use of phase unwrapping algorithms [61]). The difference between the phase unwrapping and fringe orientation unwrapping procedures is the need of multiplying by 2 the modulo $\pi$ steps, dividing the resultant unwrapped map by 2 and bringing it down to the range of LFD map, i.e., modulo $2\pi$. From the definition, in which $\beta(x,y)$ is the map of angles between vector locally normal to fringes and x axis, fringes orientation can be estimated as arctangent of the orthogonal spatial derivatives of phase function:

$$\tan(\beta(x,y)) = \frac{\partial \varphi(x,y)}{\partial x} / \frac{\partial \varphi(x,y)}{\partial y}, \quad 0 \leq \beta(x,y) < 2\pi, \tag{2}$$

$$FO(x,y) = arctan\left(\frac{\partial \varphi(x,y)}{\partial x} / \frac{\partial \varphi(x,y)}{\partial y}\right), 0 \leq FO(x,y) < \pi. \tag{3}$$

At this point it can be clearly seen that the local fringe direction map estimation is not an easy task since (1) it requires two-steps calculations and (2) the phase function needed for precise orientation calculation is encoded in the fringe pattern in the argument of cosine function, and simply it is not directly accessible in experimental reality. For that reason the orientation map cannot be calculated from the definition in the measurement reality. Instead of estimating the orthogonal spatial derivatives of phase function one can estimate the intensity gradients of the recorded fringe pattern. In the case of prefiltered fringe pattern (with uniform background, contrast and minimized noise) the intensity gradient vector has the same direction as phase gradient vector. That way the orientation map can be calculated directly from the orthogonal derivatives of the fringe pattern intensities, which is a working principle of gradient methods [39,45,57,62]. Another solution called plane fit method [31] is based on the fitting a plane polynomial (within a given window) to the gray levels of local fringes. The zero-direction derivative of the fitted plane is defined as the local fringe orientation (FO). The combined method uses both the plane-fit algorithm and gradient method [36]. Firstly the local phase gradients are approximated by plane-fitting to fringes and then those gradients are used to estimate FO. Nevertheless, the use of gradient and plane-fit algorithms requires careful adjusting of calculation window size, which is connected with the trade-off between the noise resistance (gained in the case of big window size) and higher resolution (achieved for small window size). In order to determine the local fringes orientation spin filters [26,28,29,32,33] and binary sign-maps [27,29] may be also used. Since in the experimental reality we are always dealing with the presence of noise some regularized methods [30,41,49–52] were proposed to smooth the estimated orientation maps. Other exemplifying approaches to the local fringes orientation map estimation are connected with the use of 2D energy operators [58], accumulate differences [34], Fourier transform [42], Windowed Fourier Transform [57], Principal Component Analysis [46,56] and two frame methods, e.g., optical flow [63].

However, currently proposed methods do not provide a satisfactory robustness of the fringe orientation estimation and may struggle when applying to more complex fringes (with higher local orientation variability and intensity noise). The results provided by the classical approaches strongly depend on the choice of the specific algorithm parameters. To address these issues, we propose a new, fast and robust method for fringe orientation map estimation based on convolutional neural network (CNN) called DeepOrientation. The neural networks are highly capable numerical tools for finding the relationship between their input and output signals, even though this relationship is complicated or even impossible to define analytically [64]. Additionally, the convolution is a basic operation to describe imaging process, so the CNN is an obvious choice for the task developed in this paper. CNNs were already successfully adapted in the fringe pattern analysis at different stages, i.e., conducting fringe pattern filtration [65-68], defining the optimal window for Fourier transform approach [69-71], performing phase extraction [72-76], phase unwrapping [77-82] and local fringe density map estimation [83]. Inspired by their success we decided to apply CNN to the FO map estimation. In the literature there is a neural network-based solution for fringe pattern orientation estimation [84], but it is specialized to the electronic speckle pattern interferometry (ESPI) fringe patterns. The construction of the output definition of the neural network training dataset determines that the maximum achievable accuracy is the one of the gradient method [39,62] with denoising. Considering that CNN itself is approaching the output labels with some level of error the limit defined by denoised version of gradient method not only cannot be surpassed but also reached. Since in our approach the output will be defined using the definition of the FO map from known simulated phase function the proposed DeepOrientation is a standalone and versatile solution. Additionally, in our approach input data size is preserved by DeepOrientation architecture so FO map is estimated in every pixel without reducing the analysis resolution.

The paper is structured as follows. Section 2 introduces the issue of determining fringe orientation using convolutional neural network. Section 3 contains numerical evaluation of the proposed novel neural network-based technique for the local fringe pattern orientation estimation using experimental and simulated data comparing it with the combined plane-fit/gradient method (CPFG) [36]. Section 4 contains the application of DeepOrientation to HST-based fringe pattern phase estimation comparing the obtained results with the reference TPS-based phase maps. Section 5 concludes the paper.

## 2. DeepOrientation-based fringe orientation map estimation

Facing the numerical task of transforming data input into the sought output, the solution may be found by analytic definition of the searched relationship. Naturally, this approach is connected with the full understanding of analyzed data and is mathematically solid. On the other hand, in many cases the straightforward definition of the relationship between data input and sought output may not be easy or even possible. As in the case of FO map estimation the simple definition of the relationship between the input intensity of the fringe pattern and the output orientation map is not possible since the fringe orientation by definition can be calculated from orthogonal derivatives of phase function and phase function is hidden in the intensity distribution of fringe pattern. Deep learning approach opens new possibilities for the development of algorithms solving the numerical problems one can encounter during scientific research. Deep neural networks during the supervised learning process can be taught to map the searched relationship without the need of its analytical definition. The relationship itself is defined by neural network layers parameters and algorithmic solution resolved that way works as a "black box". We can put new, unseen before by the network data instances and receive the corresponding outputs without the need of manually defining any parameter values, which is a meaningful advancement over majority of classical analytical methods. Nevertheless, because of this "black box" property neural network-based solutions raised legitimate concerns among the metrology community to use them to directly define the measurement output. For that reason, in our work, we are highlighting the use of neural network not to fully replace the mathematically sound phase estimation solutions (e.g., via HST method) but to support them. The example which is going to be discussed in this paper is the use of DeepOrientation to support the HST technique. Even if there could be some neural network-based artifacts introduced within the retrieved FO map they should not jeopardize the final HST-based phase demodulation result, as shown in our previous studies [85].

### 2.1. Definition of the training dataset

DeepOrientation network training is performed using especially tailored, simulated dataset. We decided to simulate training dataset with the uniform background modulation and without any intensity noise. That assumption was made based on the existence of robust fringe pattern filtering (denoising and detrending) algorithms [24,86-89]. Therefore, in experimental reality, well-filtered fringe patterns may be obtained. In general, the local fringe direction map is more interesting (and informative) for fringe pattern analysis and for that reason its direct estimation by neural network may seem like the most attractive solution. Nevertheless, in the case of carrier fringe pattern the fringe with the direction difference equal to $\pi$ visually appear the same, which would be confusing for the convolutional neural network during the learning process.

The process of DeepOrientation training dataset preparation is presented in Fig. 1. Using the known simulated phase function the fringe orientation map matching the simulated input fringe pattern may be calculated by the definition from orthogonal derivatives of simulated phase function (Eq. 3). The important aspect to mention at this point is the fact that in some applications (e.g., HST phase demodulation) FO map in the form of modulo $\pi$ needs to be further unwrapped to its modulo $2\pi$ form – local fringe direction map. To be able to correctly perform the unwrapping procedure the step value equal to $\pi$ must be preserved. The CNN due to the multiple convolution operations performed one after another will blur out the crucial discontinuity lines in fringe orientation map. This effect can be slightly minimized but never fully eradicated. For that reason, FO map cannot be set directly as the DeepOrientation output, because it would make the unwrapping to local fringe direction map impossible.

Now the first idea, which may come to mind is to use the known phase function orthogonal derivatives as the DeepOrientation training data output. The approach although seems very attractive is a troublesome one for the neural network learning process, because of the evenness of the cosine function. With the change of sign of the phase function the signs of its orthogonal derivatives also change while the cosines of both phase functions visually are the same. For that reason, the interpretation of the data would be confusing for neural network. Instead, another idea was formulated. The orientation angle in any point of fringe orientation map can be described in the complex form using vectorial notation. The troublesome discontinuities of the fringe orientation map can be removed by encoding it in the abovementioned way – in the form of two 2D matrices of cosine and sine functions of the orientation angle. Since the local fringe orientation (FO) map is the modulo $\pi$ indicator thus in order to use the full periodicity of sine and cosine functions the doubled fringe orientation map was encoded in their argument:

$$FO(x,y) = \frac{arg(\cos(2FO(x,y))+i\cdot\sin(2FO(x,y)))}{2}. \tag{4}$$

Thus, two maps of $\cos(2FO)$ and $\sin(2FO)$ define the neural network output. DeepOrientation inputs (I(x,y), see exemplary fringe patterns in Fig. 1) were generated as in (Eq. 5):

$$I(x,y) = \cos(\varphi_{obj}(x,y) + \varphi_{carrier}(x,y)),  \quad (5)$$

where $\varphi_{obj}(x,y)$ is the object phase function simulated as a sum of dozens (up to 50) 2D Gaussian kernels, each one with random standard deviation and $(x,y)$ location, $\varphi_{carrier}(x,y)$ is the factor that generates carrier fringes with random orientation ($\theta$) and period ($T$):

$$\varphi_{carrier}(x,y) = x\frac{\cos(\theta)\cdot 2\pi}{T} + y\frac{\sin(\theta)\cdot 2\pi}{T}. \quad (6)$$

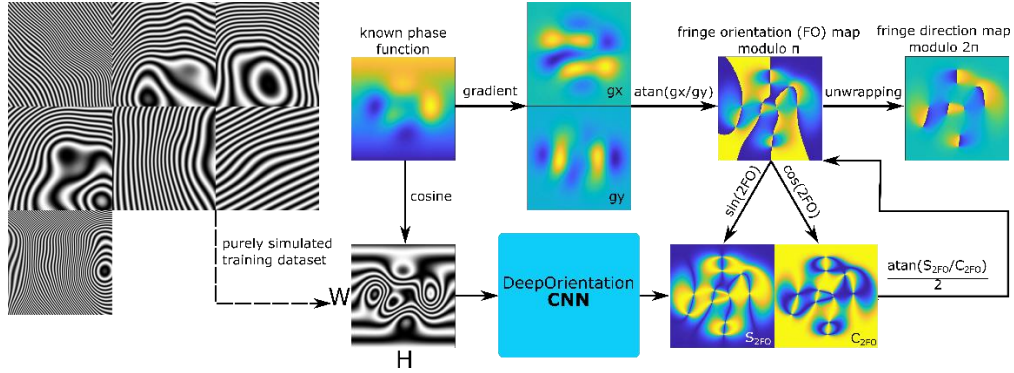

Fig. 1. Training and working principle of DeepOrientation convolutional neural network.

## 2.2. Proposed network architecture

The DeepOrientation network architecture schematically presented in Fig. 2 was inspired by the work [72] and already successfully adaptation to somewhat similarly challenging task of local fringe density map estimation [84]. DeepOrientation data input is a grayscale image, in other words one-channel 2D matrix. The network architecture is built by convolutional layers and residual blocks. It is divided into different paths where the input image dimensionality is changed by the maxpooling layers. By the end of each path the results are upsampled to match the input image height and width and then results from all paths are concatenated to define the input for final convolutional layer. The last convolutional layer defines the DeepOrientation data output to have two-channels with height and width matched to the input image. During further analysis two parameters will be adjusted to optimize the network architecture and adapt it to the specific task of FO map estimation: number of paths and number of filters in convolutional layers (including those building the residual blocks). Increasement of those two parameters makes the network architecture more complex. Because in our approach the training dataset is simple and was used for grasping the general relationship between the fringe pattern and underlying orientation map it was crucial to prevent the network from overfitting to the trained data. In order to do that the residual blocks with skip connections were chosen.

Training process was performed on a training dataset containing 2400 512x512 px images. During the training, the mini batch size was equal to 1 and initial learning rate was $10^{-4}$. Learning rate was updated each 5 epochs and reduced by the factor of 5 to help the loss function get out of local minima. The ADAM optimizer was used as a solver for training network and the mean-squared-error function was used as the loss function. Learning process lasted for 30 epochs, which was enough for the networks to train since no significant further decrease of loss function was observed afterwards. Networks were trained on a computer with AMD Ryzen 9 5900X 12-Core 3.70 GHz processor and NVIDIA GeForce RTX 3080 graphics card with 12 GB of memory, that allowed to train a single network in the time between 200 and 2000 minutes, depending on the architecture complexity. It is worth to highlight that this time-consuming training process needs to be performed only once for a given architecture. After the training, networks can reconstruct the orientation of a 512x512 px fringe pattern image in less than a second. Considering available memory on our GPU, networks with bigger number of filters and paths could only be trained with a mini batch size equal to 1. To keep the learning process consistent among all networks we used the same mini batch size for all trainings.

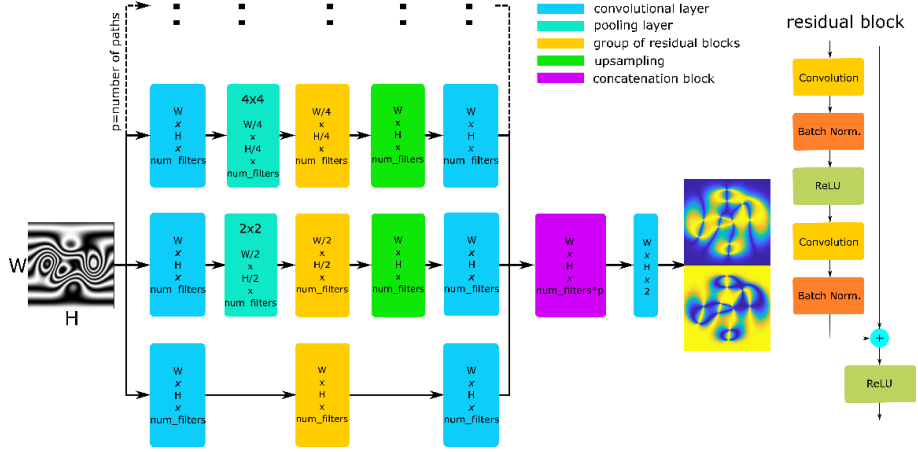

Fig. 2. Scheme of the developed DeepOrientation convolutional neural network architecture.

*2.3. Influence of the neural network architecture complexity on the learning accuracy*

In a pursuit to find the optimal neural network architecture for DeepOrientation two parameters were considered – number of paths with different downsampling and number of filters in convolutional layers. Increase of each of those parameters caused the increase of the neural network architecture complexity. In total 24 different configurations were tested with the number of paths varying from 2 to 5 and the number of filters (per path) varying from 30 to 130 with the step of 20, which as can be seen in Fig. 3. Our study allowed to understand general relationships between the network complexity, accuracy and calculation time. The performance of developed neural networks was tested with the use of two datasets with different definition of data instances. The dataset called validation set (600 512x512 px images) was used to test the performance of neural networks during training and is of the same origin as training dataset. Second dataset called test set is also based on simulations (Eq. 5), but the object phase functions included there were simulated in a completely different manner in order to validate the generalization ability of proposed DeepOrientation network. Test set consisted of 5 different $\varphi_{obj}(x, y)$ functions: (1) a 2D function with 3 maxima and 2 minima (simulated using MATLAB 'peaks' function obtained by translating and scaling Gaussian distributions), (2) a group of 5 HeLa cells with shapes that were close to spherical, (3) a group of 2 HeLa cells with oblong shapes, (4) a blurred binary mask of human hand and (5) a group of 23 grains of rice. For each of those functions, there were generated a 140 fringe patterns with different carrier fringes period and orientation, and with different fringes curvature (introduced by changing the dynamic range of the $\varphi_{obj}(x, y)$ function). Exemplary test set image may be seen in Fig. 4(a).

Choosing the optimal neural network architecture for the specific task of local fringe pattern orientation map estimation is a complex issue, which needs to be carefully analyzed. The training strategy picked for DeepOrientation was based on the assumption of the simple simulated training dataset (without noise, background and amplitude modulation). Subsequently trained network is supposed to work for a wide range of fringe pattern characteristics, where phase function may not necessarily be describable the same way as phase functions included in training dataset. For that reason, we need to be especially careful to not introduce overfitting in wider sense that during the standard neural network training. Even if the neural network is not overfitted in the sense of being able to successfully analyze the data, which was introduced during the training, it can still 'overfit' assuming that all data outside the training dataset is of the same characteristics and origin (shape of fringes, optical measurement method used and studied object type). In other words, we want to find the solution leading to the estimation of the FO map from the cosine pattern, but without the strong restriction that the phase function needs to be describable the way proposed in training dataset simulation.

In Fig. 3 the results of the performance analysis for different levels of neural network architecture complexity are presented. Looking at the curves in Fig. 3(a) estimated with the use of a validation dataset one can notice that with the increase of filters number adding the extra paths does not influence the results accuracy. For the filters number greater than 90 all neural networks achieved similar accuracy regardless the number of paths. Nevertheless, it needs to be highlighted that with the increase of the architecture complexity the neural network ability to fit to the training dataset increases. As it was just discussed with the chosen training strategy, we do not want to fit perfectly only to the training dataset. Observing the Fig. 3(a) curves estimated for the test dataset the first aspect one can notice is the increase of the RMSE value, which is perfectly understandable since the origin of the test data is different than training dataset (as it would be in different experimental realities – setups, objects) and some of the data included in the test dataset featured higher phase gradients than the validation dataset. It can be clearly seen in error maps presented in Fig. 4, in which the highest errors are visible around the edges of HeLa cells where phase gradients are the highest. Nevertheless, the error values are still on the reasonable level especially considering the main planned application of DeepOrientation network, which is to support HST-based phase estimation. Despite the obvious change in the error values the test curves shape also changed in comparison with the validation curves. The minimum RMSE was achieved for the neural network with two paths and 110 filters, therefore this configuration was chosen for the final DeepOrientation architecture. Two paths architecture limits the complexity of possible neural network input-output relationship preventing too strong fitting to the training dataset structure, while 110 filters grant that the network architecture is complex enough to capture the general relationship (since for that number of filters there was no noticeable error difference obtained on validation dataset for different number of paths).

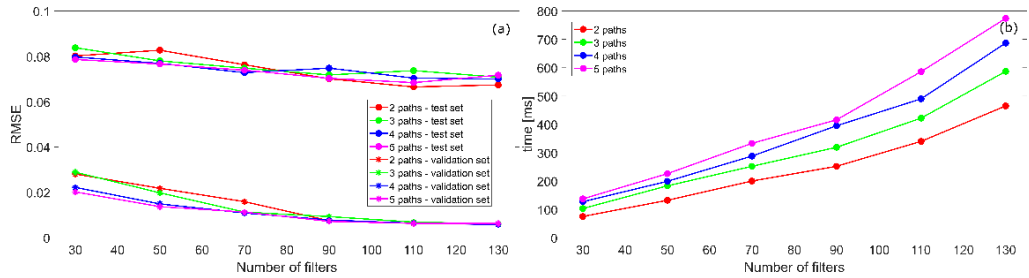

Fig. 3. The performance of neural network architectures with different level of complexity trained to estimate fringe pattern orientation maps: (a) the mean RMSE values calculated on validation and test datasets and (b) calculation time of single data instance.

The detailed error analysis of the neural networks' outputs generated by exemplary fringe pattern from test dataset is presented in Fig. 4. One can notice that in general with the increase of neural network complexity, either implemented by increasing the number of filters or paths, the presented error maps become darker, which indicates that mean error value is decreasing. On the other hand, error map estimated for DeepOrientation architecture (i.e., 110 filters and 2 paths) has lower errors in the regions of high phase gradient (see circular cell fragment visible at the bottom). Presented error maps are estimated as absolute value of difference between the sine of known, ground truth doubled FO map and sine output of neural networks. We demonstrate the results connected only with sine output, because maps estimated for cosine output are complementary and do not contribute new information to the discussion.

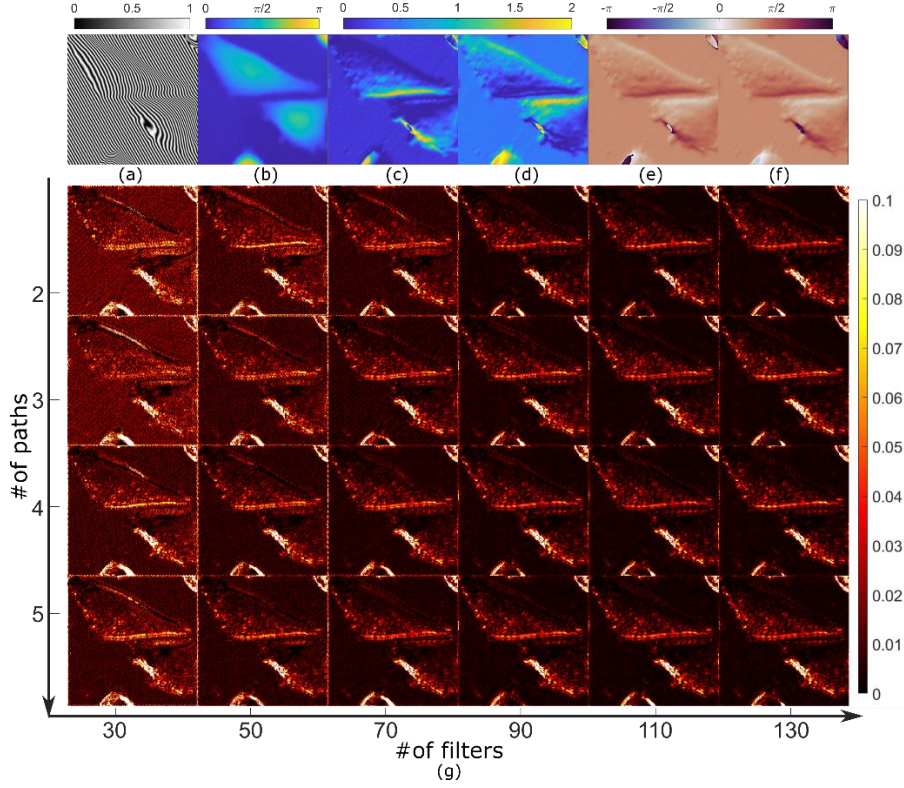

Fig. 4. Error analysis of developed neural networks. (a) Analyzed fringe pattern from test dataset; (b) underlying phase function; ground truth outputs of DeepOrientation neural network: (c) sine and (d) cosine of 2FO; (e) ground truth FO map and (f) its unwrapped version: local fringe direction map; (g) error maps of sin(2FO) output for all analyzed neural network architectures.

Additional factor, which was considered while choosing the DeepOrientation network architecture was calculation time. From the algorithm's user perspective, one of the most important information is to know how long it would take to process their data. For that reason, in Fig. 3(b) the time needed for the calculations of the single data instance was presented. Reported calculation times were estimated with the use of typical computing unit represented by personal laptop (Intel Core i7-7700HQ 2.80 GHz processor and NVIDIA GeForce GTX 1060 graphics card). Obtained values confirm that unnecessary augmentation of neural network architecture complexity is undesirable.

## 3. Numerical evaluation of DeepOrientation

The analysis comparing our proposed DeepOrientation approach with classical CPFG method [35] using simulated data is presented in Fig. 5 and using experimental data in Figs. 6 and 7. Since the local orientation maps consist of the angle information, in order to preserve its periodic nature, we introduced the orientation error (OE) as:

$$OE = \sqrt{\frac{1}{N_x N_y - 1} \sum_{x=1}^{N_x} \sum_{y=1}^{N_y} \left[\sin\left(FO(x,y) - FO_{ref}(x,y)\right) - \mu\right]^2}, \quad (7)$$

where $N_x$ and $N_y$ are image size, $FO_{ref}(x,y)$ is a reference local fringe orientation map and $\mu$ is mean of $\sin\left(FO(x,y) - FO_{ref}(x,y)\right)$. In other words orientation error may be considered as modified RMSE, where the straightforward difference between retrieved map and its ground truth was replaced by the sine of that difference. The orientation error converges to 0 if the

$FO(x, y) - FO_{ref}(x, y)$ is equal to an integer multiple of $\pi$, which is a desirable feature since orientation map is in the form of modulo $\pi$.

### 3.1. Comparison of DeepOrientation with classical approach on simulated data

The fringe pattern series used for analysis in Fig. 5 were simulated according to the (Eq. 5) and (Eq. 6), where T=14, $\theta = 0$ and $\varphi_{obj}(x, y)$ is described by Matlab peaks function with dynamic range controlled by multiplication by $a$ coefficient varying from 0 to 10.

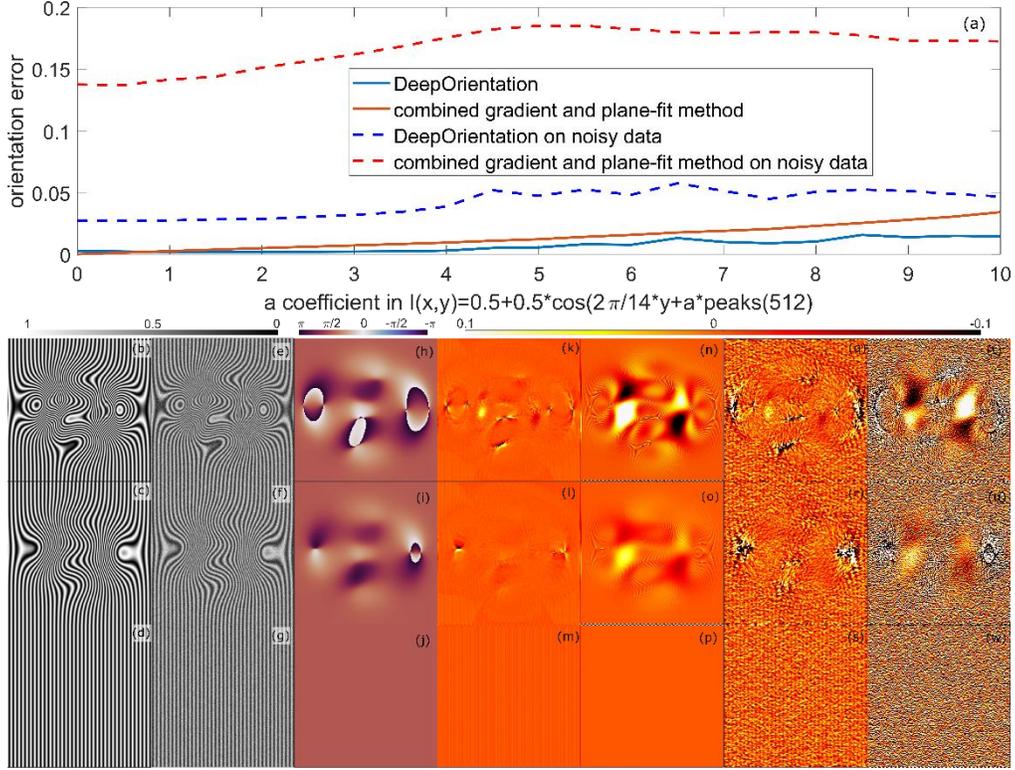

Fig. 5. Comparison of the performance of DeepOrientation approach and classical one (CPFG [35]) using simulated fringe patterns. (a) The orientation errors of both methods calculated for different levels of phase modulation, (b), (c), (d) exemplary fringe patterns with high (a=10), medium (a=5) and low (a=0) phase modulation, respectively, (e), (f), (g) noisy versions of fringe patterns from (b), (c), (d), respectively, (h), (i), (j) the ground truth FO maps for (b), (c), (d), respectively, orientation error maps estimated by (k), (l), (m) DeepOrientation and (n), (o), (p) CPFG method for (b), (c), (d), respectively and orientation error maps estimated by (q), (r), (s) DeepOrientation and (t), (u), (w) CPFG method for (e), (f), (g), respectively.

In the case of CPFG method the parameter, which needs to be set is the size of the window in which the orientation angle will be estimated. The smaller the window size, the greater the accuracy of local orientation estimation. Nevertheless, small window size is not immune to the noise presence and for that reason in many cases it is recommended to set the bigger window sizes. Since the DeepOrientation works on prefiltered data in order to provide a fair comparison between two algorithms throughout the paper we are going to use the prefiltered data also for the classical approach. This can be considered as novel modification of CPFG aimed at its automation (no need for tailoring the window size) and increasement of robustness via unsupervised variational image decomposition (uVID) fringe prefiltering [87] and HST-based fringes normalization [23]. For that reason the window size can be chosen arbitrarily small so the value 2 was used in all presented cases. We have tested the CPFG accuracy using different window sizes and in majority of cases (if the denoising was correctly performed) the window size equal to 2 provided the best results. It can be seen that for low level of phase modulation ($a < 1$) CPFG method provides higher accuracy of the retrieved local orientation maps. As it is shown in Figs. 5(d), 5(j), 5(m) and 5(p) DeepOrientation-based results have a small fringe-like error, while for such simple cases and perfectly fitted window size classical CPFG approach provides error-free result. Nevertheless, with the increase of phase modulation level (and therefore complication of the fringe pattern shape itself) the predominance of DeepOrientation approach is clearly visible. It is also worth to mention that the orientation errors values presented in Fig. 5(a) were calculated after neglecting the border effects, which are obvious in the case of CPFG method even in the case of small window size. Additionally, DeepOrientation is more resistant to noise errors than CPFG method, which can be clearly see in Fig. 5(a). If there is noise present as in the case of Figs. 5(e)-5(g), where the Gaussian noise of std=0.1 was added to the data from Figs. 5(a)-5(d), DeepOrientation provides smoother orientation maps than CPFG method with smallest window size. The CPFG method error could be minimized by adjusting the window size and match the DeepOrientation accuracy, which shows how troublesome and crucial parameter's adjusting could be for a classical method.

### 3.2. Experimental verification of the accuracy of DeepOrientation-based local fringe orientation map estimation

The performance of proposed DeepOrientation solution was also tested using the experimentally recorded fringe patterns and compared with classical, well-developed solution represented by CPFG method [35]. All analyzed experimentally recorded data was prefiltered with the use of uVID [87] (where the noise part of the decomposition is estimated with the use of BM3D) and normalized in 0-1 range with the use of HST approach [23] before calculating the orientation map either with the use of the DeepOrientation or the CPFG. The first real-life example we have chosen contains complicated, low frequency fringe patterns recorded during the temporal phase shifting (TPS) study of glass plate in Twyman-Green interferometer; fringe patterns are presented in Figs. 6(a)-6(e). Having the complete TPS series we were able to precisely calculate the reference phase map since the TPS algorithm (as the multi-frame fringe pattern analysis algorithm) is the most accurate phase demodulation method, especially in the case of sparse closed fringes. Using this reference phase map and the definition of the FO map (Eq. 3) the reference FO map was calculated and can be seen in Fig. 6(p). One can notice that presented FO map is very noisy. It is due to the fact that 5-frames TPS algorithm is not fully resistant to the presence of noise and unfiltered intensity noise is transferred to the retrieved phase map. The noise effect is further amplified in the case of FO map estimation because of the needed numerical gradients calculation. For that reason, the denoised (using block-matching 3D denoising (BM3D) algorithm [86] on every analyzed intensity frame) version of estimated FO map is presented in Fig. 6(r) and that map will be further deployed as the reference for estimating the orientation error values. As it can be clearly seen analyzing the orientation error values shown in Table 1 in all cases (for all single-shot fringe pattern frames) the DeepOrientation provided better results than the CPFG method. Additionally, comparing the DeepOrientation results (Fig. 6(f)-6(j)) and the classical approach results (Fig. 6(k)-6(o)) the first ones have better preserved edges (on the modulo π steps), which is especially important as one of the planned use of DeepOrientation is a support for single-fringe-pattern HST-base phase estimation. The reason is that FO map unwrapping procedure [61] needs a clear, well-preserved steps values to provide a correct unwrapping.

To evaluate DeepOrientation on the biological data, Fig. 7, we collected 10, phase-shifted interferograms of a group of HeLa cells on a Linnik interferometer [90]. Similarly as above, we used the TPS method aided with BM3D denoising [86] to reconstruct cells phase, which was then used to obtain reference FO map, Fig. 7(b). Next, we prefiltered one of the collected interferograms with the uVID algorithm and obtained orientation maps with DeepOrientation, Fig. 7(c), and CPFG, Fig. 7(d), algorithms. Both methods returned results that were close to the reference map with orientation error equal to 0.1843 for Fig. 7(c), 0.1925 for Fig. 7(d), 0.1191 for Fig. 7(g), 0.1579 for Fig. 7(h), 0.1672 for Fig. 7(k) and 0.1916 for Fig. 7(l). However, as can be observed on a zoomed parts of the reconstructed maps (Figs. 7(f)-7(h) and 7(j)-(l)), the CPFG reconstruction has some unexpected orientation jumps along the fringe profile, whereas DeepOrientation reconstruction is much smoother. This indicates that DeepOrientation is more robust to fringe patterns being transferred to the orientation map than the CPFG method.

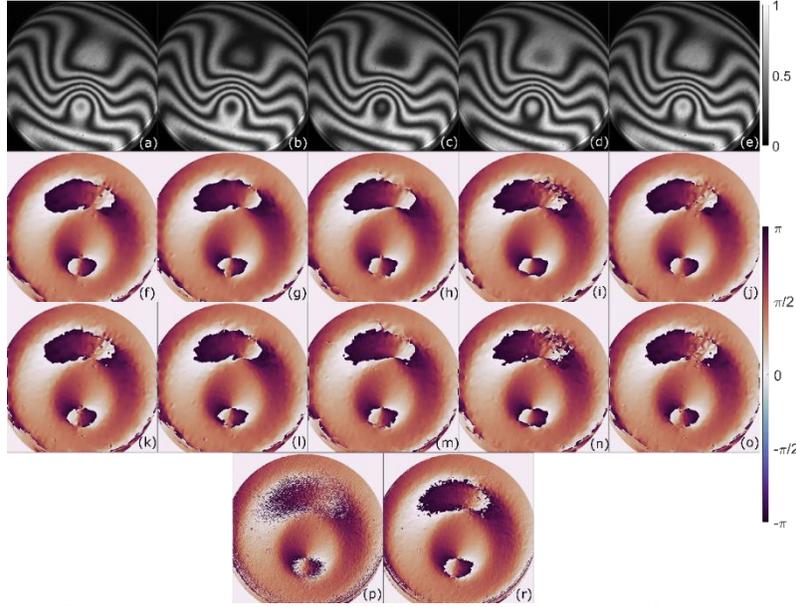

Fig. 6. Experimentally recorded TPS series of interferograms with phase shift equal to π/2: (a-e) subsequent interferograms, (f-j) FO maps calculated by DeepOrientation, (k-o) FO maps calculated by CPFG method [35], (p) FO map calculated from TPS estimated phase function, (r) FO map calculated from TPS estimated phase function with BM3D denoising.

Table 1. Numerical analysis of the accuracy of estimated results from Fig. 6.

| image number<br>orientation error | 1 | 2 | 3 | 4 | 5 |
|---|---|---|---|---|---|
| DeepOrientation | 0.1627 | 0.1562 | 0.1722 | 0.1802 | 0.1764 |
| CPFG | 0.1684 | 0.1628 | 0.1764 | 0.1893 | 0.1806 |

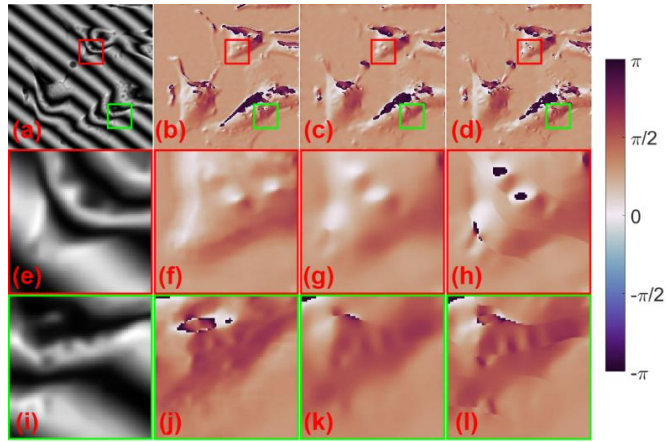

Fig. 7. One of the recorded fringe pattern images of the HeLa cells (a), reference local orientation map obtained from the TPS retrieved phase (b), reconstructed local orientation maps from the single prefiltered fringe pattern image with the use of DeepOrientation (c) and CPFG (d) methods. Zoomed parts of the (a)-(d) images inside red (e)-(h) and green (i)-(l) boxes.

## 4. The influence of DeepOrientation onto the accuracy of the HST-based single-shot fringe-pattern phase estimation

The one of possible applications of DeepOrientation is guiding the phase demodulation process for Hilbert spiral transform [23]. As a result of HST the quadrature fringe function is obtained with phase shift equal to $0.5\pi$ introduced between input $s(x,y)$ and output $s_H(x,y)$. The important thing worth to emphasize is that HST needs a zero mean value signal as an input, therefore successful fringe pattern background removal is of the essence. Additionally, it is recommended to minimize the intensity noise for the retrieved phase map quality improvement. Therefore, the HST input signal can be described as:

$$s(x,y) = b(x,y)cos(\varphi(x,y)), \qquad (8)$$

and then output signal follows as:

$$s_H(x,y) = -b(x,y)sin(\varphi(x,y)). \qquad (9)$$

Finally, the phase function can be calculated as:

$$\varphi(x,y) = \tan^{-1}\left(\frac{s_H(x,y)}{s(x,y)}\right). \qquad (10)$$

Using the HST nomenclature [23] the quadrature function can be described as:

$$s_H(x,y) = -iexp[-\beta(x,y)]F^{-1}\{S(u,v)F\{s(x,y)\}\}, \qquad (11)$$

where $F$ denotes Fourier transform, $F^{-1}$ denotes inverse Fourier transform, $S(u,v)$ is spiral phase function defined in spatial frequencies $(u,v)$ domain and $\beta(x,y)$ is LFD map. The LFD map is instrumental as it guides the phase demodulation process. It is especially important in the case of very complicated, overlapping fringe pattern spectrum. Correct LFD map helps to avoid sign ambiguity errors in closed (concentric) fringe pattern phase demodulation.

We would like to highlight that the DeepOrientation is not employed here to directly determine the phase function, the outcome of the optical measurement. The use of neural network to replace the mathematically rigorous phase estimation algorithmic derivation may raise legitimate metrological concerns. For that reason, in our work the HST phase calculations are only supported by DeepOrientation neural network, which constitutes our novel approach. DeepOrientation allows the estimation of the FO map, which afterwards is unwrapped [61] to local fringe direction map and used to guide the HST-driven phase estimation process.

To prove that DeepOrientation is a valuable tool in terms of aiding HST algorithm with phase retrieval, Fig. 8, we collected a 3 data series consisting of 5 phase-shifted interferograms of HeLa cells, exemplifying one shown in Fig. 8(a), LSEC cells, exemplifying one presented in Fig. 8(e), and phase test target, exemplifying one depicted in Fig. 8(i).

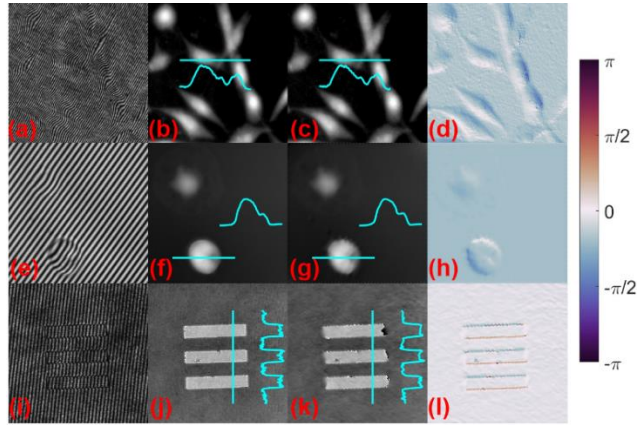

Fig. 8. One of the recorded interferograms of HeLa cells (a), LSEC cells (e) and phase test target (i). Reconstructed reference phase maps from TPS data (b),(f),(j), reconstructed phase maps by HST (c),(g),(k) and reconstructed local fringe direction maps with the use of DeepOrientation algorithm (d),(h),(l). Phase maps are given in range 0-18 (b),(c), 0-6 (f),(g) and 0-8 (j),(k) rad.

Next, from those interferograms we retrieved the reference phase maps with the use of TPS algorithm aided with BM3D method, Figs. 8(b), 8(f) and 8(j), respectively. After that, from each data series, we filtered a single interferogram with the uVID algorithm [87], which was then provided to DeepOrientation to reconstruct local fringe pattern orientation map. Those maps were then unwrapped with the use of phase unwrapping algorithm presented in [61] to obtain local fringe direction maps, Figs. 8(d), 8(h) and 8(l). At the end, filtered fringe patterns along with obtained fringe direction maps were supplied to the HST algorithm to reconstruct phase maps, Figs. 8(c), 8(g) and 8(k). One can noticed that HST-based results estimated with the use of single-frame approach compare favorably with the highly accurate multi-frame approach. To be exact the RMSE for HST-based results is equal to 0.0132 rad for Fig. 8(c), 0.0132 rad for Fig. 8(g) and 0.0521 rad for Fig. 8(k). This fact corroborated DeepOrientation guided HST for quantitative phase imaging of living biosamples and challenging technical objects.

## 5. Conclusions

In this paper, we have proposed an accurate, robust, and fast numerical solution for the local fringe orientation map estimation called DeepOrientation based on neural networks and deep learning. The fringe patterns themselves are the example of ideal data for neural network training process. Even if the underlying phase function varies drastically between different measurements, fringe patterns generally have a similar structure as most of them can be described by a spatially self-similar cosine function. That makes the learning process easier, and we have shown that reliable network parameters can be learned based on a relatively small training dataset, not highly diverse in the meaning of phase function characteristic. The DeepOrientation works well even for the data, where underlying phase function significantly differs from the ones included in the training dataset, due to general self-similarity of all fringe patterns. The validity and effectiveness of the DeepOrientation were corroborated both on simulated and experimental data and compared favorably with the classical approach. It should be noted that once the DeepOrientation training is finished, the parameters do not need to be further adjusted, as the trained network generalizes sufficiently. We have provided a solution, which was tested on a wide range of fringe pattern and can be used on the new fringe data instances without additional adjusting or retraining. Additionally, DeepOrientation fills the gap in the search for increasingly accurate fringe pattern analysis tools. As it was shown it can be successfully employed for guidance of single-shot phase demodulation process in Hilbert spiral transform and there are plenty of other possible applications for it [26-59].

**Funding**

This work has been partially funded by the National Science Center Poland (OPUS 2020/37/B/ST7/03629 and PRELUDIUM 2021/41/N/ST7/04057). Studies were funded by FOTECH-1 project granted by Warsaw University of Technology under the program Excellence Initiative: Research University (ID-UB). MC work was supported by the Foundation for Polish Science (FNP) and by the Polish National Agency for Academic Exchange under the Iwanowska programme.

**Disclosures**

The author declares no conflicts of interest.

**Data Availability.**

Data may be obtained from the authors upon reasonable request. Trained DeepOrientation model is made freely available in Ref. [91].

## References

1. J. Schwider, "Advanced evaluation techniques in interferometry," in Progress in Optics, E. Wolf, ed., (Elsevier, 1990).
2. D. W. Robinson and G. Reid, *Interferogram Analysis: Digital Fringe Pattern Measurement* (Institute of Physics Publishing, 1993).
3. D. Malacara, M. Servin, and Z. Malacara, *Interferogram Analysis for Optical Testing* (Marcel Dekker, 1998).
4. M. K. Kim, *Digital Holographic Microscopy: Principles, Techniques, and Applications* (Springer-Verlag, 2011).
5. B. Kemper and G. von Bally, "Digital holographic microscopy for live cell applications and technical inspection," Appl. Opt. **47**(4), A52-A61 (2008).
6. M. K. Kim, "Principles and techniques of digital holographic microscopy," SPIE Rev. **1**(1), 018005 (2010).
7. S. S. Gorthi and P. Rastogi, "Fringe projection techniques: Whither we are?" Opt. Lasers Eng. **48**(2), 133–140 (2010).
8. R. Sitnik, "Four-dimensional measurement by a single-frame structured light method," Appl. Opt. **48**(18), 3344 (2009).
9. K. Patorski and M. Kujawińska, *Handbook of the Moirè Fringe Technique* (Elsevier, 1993).
10. M. Takeda, H. Ina and S. Kobayashi, "Fourier-transform method of fringe-pattern analysis for computer-based topography and interferometry," J. Opt. Soc. Am. **72**(1), 156–60 (1982).
11. Q. Kemao, Windowed Fringe Pattern Analysis (Bellingham, WA: SPIE, 2013)
12. K. Pokorski and K. Patorski, „Processing and phase analysis of fringe patterns with contrast reversals," Opt. Express **21**(19), 22596–609 (2013).
13. X. Guo, H. Zhao and X. Wang, "Phase retrieval from a single fringe pattern by using empirical wavelet transform," Meas. Sci. Technol. **26**(9), 095208 (2015).
14. M. Pirga and M. Kujawińska, "Two directional spatial carrier phase-shifting method for analysis of crossed and closed fringe patterns" Opt. Eng. **34**(8), 2459–2466 (1995).
15. L. Kai and Q. Kemao, "Improved generalized regularized phase tracker for demodulation of a single fringe pattern," Opt. Express **21**(20), 24385–24397 (2013).
16. Y. Baek, K. Lee, S. Shin, and Y. Park, "Kramers–Kronig holographic imaging for high-space-bandwidth product," Optica **6**(1), 45-51 (2019).
17. Y. Tounsi, S. Zada, D. Muhire, A. Siari, and A. Nassim, „ Estimation of phase derivative from a single fringe pattern using Riesz transforms," Opt. Eng. **56**(11), 111706 (2017).
18. Y. Tounsi, M. Kumar, A. Siari, F. Mendoza-Santoyo, A. Nassim, and O. Matoba, „Digital four-step phase-shifting technique from a single fringe pattern using Riesz transform," Opt. Lett. **44**(14), 3434-3437 (2019).
19. T. Ikeda, G. Popescu, R. R. Dasari, and M. S. Feld, "Hilbert phase microscopy for investigating fast dynamics in transparent systems," Opt. Lett. **30**(10), 1165-1167 (2005).
20. L. Xue, J. Lai, S. Wang, and Z. Li, "Single-shot slightly-off-axis interferometry based Hilbert phase microscopy of red blood cells," Biomed. Opt. Express **2**(4), 987-995 (2011).
21. M. Trusiak, V. Mico, J. Garcia, and K. Patorski, „Quantitative phase imaging by single-shot Hilbert–Huang phase microscopy," Opt. Express **41**(18), 4344-4347 (2016).
22. N. T. Shaked, Y. Zhu, M. T. Rinehart, and A. Wax, "Two-step-only phase-shifting interferometry with optimized detector bandwidth for microscopy of live cells," Opt. Express **17**(18), 15585-15591 (2009).
23. K. G. Larkin, D. J. Bone, and M. A. Oldfield, "Natural demodulation of two-dimensional fringe patterns. I. General background of the spiral phase quadrature transform," J. Opt. Soc. Am. A **18**(8), 1862–1870 (2001).
24. M. Trusiak, K. Patorski and M. Wielgus, "Adaptive enhancement of optical fringe patterns by selective reconstruction using FABEMD algorithm and Hilbert spiral transform," Opt. Express **20**(21), 23463–23479 (2012).
25. M. Trusiak, M. Cywinska, V. Micó, J. Á. Picazo-Bueno, C. Zuo, P. Zdańkowski and K. Patorski, "Variational Hilbert Quantitative Phase Imaging," Sci Rep **10**, 13955 (2020)
26. Q. Yu, X. Liu, and K. Andresen, "New spin filters for interferometric fringe patterns and grating patterns," Appl. Opt. **33**(17), 3705–3711 (1994).
27. Q. Yu and K. Andresen, "Fringe-orientation maps and fringe skeleton extraction by the two-dimensional derivative-sign binary-fringe method," Appl. Opt. **33**(29), 6873–6878 (1994).
28. Q. Yu, K. Andresen, W. Osten, and W. Jueptner, "Noise-free normalized fringe patterns and local pixel transforms for strain extraction," Appl. Opt. **35**(20), 3783–3790 (1996).
29. Q. Yu, X. Liu, and X. Sun, "Generalized spin filtering and an improved derivative-sign binary image method for the extraction of fringe skeletons," Appl. Opt. **37**(20), 4504–4509 (1998).
30. J. L. Marroquin, R. Rodriguez-Vera, and M. Servin, "Local phase from local orientation by solution of a sequence of linear systems," J. Opt. Soc. Am. A **15**(6), 1536–1544 (1998).
31. Q. Yu, X. Sun, X. Liu, and Z. Qiu, "Spin filtering with curve windows for interferometric fringe patterns," Appl. Opt. **41**(14), 2650–2654 (2002).
32. Q. Yu, S. Fu, X. Liu, X. Yang, and X. Sun, "Single-phase-step method with contoured correlation fringe patterns for ESPI," Opt. Express **12**(20), 4980–4985 (2004).
33. Q. Yu, X. Yang, S. Fu, and X. Sun, "Two improved algorithms with which to obtain contoured windows for fringe patterns generated by electronic speckle-pattern interferometry," Appl. Opt. **44**(33), 7050–7054 (2005).

34. X. Yang, Q. Yu, and S. Fu, "An algorithm for estimating both fringe orientation and fringe density," Opt. Commun. **274**(2), 286–292 (2007).
35. X. Yang, Q. Yu, and S. Fu, "A combined method for obtaining fringe orientations of ESPI," Opt. Commun. **273**(1), 60–66 (2007).
36. X. Yang, Q. Yu, and S. Fu, "Determination of skeleton and sign map for phase obtaining from a single ESPI image," Opt. Commun. **282**(12), 2301–2306 (2009).
37. F. Zhang, W. Liu, J. Wang, Y. Zhu, and L. Xia, "Anisotropic partial differential equation noise-reduction algorithm based on fringe feature for ESPI," Opt. Commun. **282**(12), 2318–2326 (2009).
38. A. M. Siddiolo and L. D'Acquisto, "A direction/orientation- based method for shape measurement by shadow Moire," IEEE Trans. Instrum. Meas. **57**(4), 843–849 (2008).
39. C. Tang, L. Han, H. Ren, D. Zhou, Y. Chang, X. Wang, and X. Cui, "Second-order oriented partial-differential equations for denoising in electronic-speckle-pattern interferometry fringes," Opt. Lett. **33**(19), 2179–2181 (2008).
40. H. Wang, Q. Kemao, W. Gao, F. Lin, and H. S. Seah, "Fringe pattern denoising using coherence-enhancing diffusion," Opt. Lett. **34**(8), 1141–1143 (2009).
41. J. Villa, J. A. Quiroga, and I. De la Rosa, "Regularized quadratic cost function for oriented fringe-pattern filtering," Opt. Lett. **34**(11), 1741–1743 (2009).
42. C. Tang, Z. Wang, L. Wang, J. Wu, T. Gao, and S. Yan, "Estimation of fringe orientation for optical fringe patterns with poor quality based on Fourier transform," Appl. Opt. **49**(4), 554–561 (2010).
43. H. Wang, Q. Kemao, R. Liang, H. Wang, M. Zhao, and X. He, "Oriented boundary padding for iterative and oriented fringe pattern denoising techniques," Sig. Proc. **102**(9), 112–121 (2014).
44. D. Zhang, M. Ma, and D. D. Arola, "Fringe skeletonizing using an improved derivative sign binary method," Opt. Lasers Eng. **37**(1), 51–62 (2002).
45. C. Tang, W. Lu, Y. Cai, L. Han, and G. Wang, "Nearly preprocessing-free method for skeletonization of grayscale electronic speckle pattern interferometry fringe patterns via partial differential equations," Opt. Lett. **33**(2), 183–185 (2008).
46. Z. Zhang and H. Guo, "Principal-vector-directed fringe-tracking technique," Appl. Opt. **53**(31), 7381–7393 (2014).
47. Q. Kemao and S. Hock Soon, "Sequential demodulation of a single fringe pattern guided by local frequencies," Opt. Lett. **32**(2), 127–129 (2007).
48. H. Wang and Q. Kemao, "Frequency guided methods for demodulation of a single fringe pattern," Opt. Express **17**(17), 15118–15127 (2009).
49. J. A. Quiroga, M. Servin, J. L. Marroquin, and D. Crespo, "Estimation of the orientation term of the general quadrature transform from a single n-dimensional fringe pattern," J. Opt. Soc. Am. A **22**(3), 439–444 (2005).
50. O. S. Dalmau-Cedeño, M. Rivera, and R. Legarda-Saenz, "Fast phase recovery from a single closed-fringe pattern," J. Opt. Soc. Am. A **25**(6), 1361–1370 (2008).
51. M. Rivera, "Robust phase demodulation of interferograms with open or closed fringes," J. Opt. Soc. Am. A **22**(6), 1170–1175 (2005).
52. J. Villa, J. A. Quiroga, M. Servin, J. C. Estrada, and I. de la Rosa, "N-dimensional regularized fringe direction estimator," Opt. Express **18**(16), 16567–16572 (2010).
53. S. Dehaeck, Y. Tsoumpas, and P. Colinet, "Analyzing closed-fringe images using two-dimensional Fan wavelets," Appl. Opt. **54**(10), 2939–2952 (2015).
54. E. Robin, V. Valle, and F. Brémand, "Phase demodulation method from a single fringe pattern based on correlation with a polynomial form," Appl. Opt. **44**(34), 7261–7269 (2005).
55. B. Deepan, C. Quan, and C. J. Tay, "Determination of phase derivatives from a single fringe pattern using Teager-Hilbert-Huang transform," Opt. Commun. **359**, 162–170 (2016).
56. Z. Zhang and H. Guo, "Fringe phase extraction using windowed Fourier transform guided by principal component analysis," Appl. Opt. **52**(27), 6804–6812 (2013).
57. H. Wang and Q. Kemao, "Quality-guided orientation unwrapping for fringe direction estimation," Appl. Opt. **51**(4), 413–421 (2012).
58. K. Larkin, "Uniform estimation of orientation using local and nonlocal 2-D energy operators," Opt. Express **13**(20), 8097–8121 (2005).
59. J. A. Quiroga, M. Servin, and F. Cuevas, "Modulo 2π fringe orientation angle estimation by phase unwrapping with a regularized phase tracking algorithm," J. Opt. Soc. Am. A **19**(8), 1524–1531 (2002).
60. B. Jahne, *Practical Handbook on Image Processing for Scientific Applications* (CRC, 1997).
61. M. A. Herráez, D. R. Burton, M. J. Lalor, and M. A. Gdeisat, "Fast two-dimensional phase-unwrapping algorithm based on sorting by reliability following a noncontinuous path," Appl. Opt. **41**(35), 7437-7444 (2002).
62. X. Zhou, J. P. Baird, and J. F. Arnold, "Fringe-orientation estimation by use of a Gaussian gradient filter and neighboring-direction averaging," Appl. Opt. **38**(5), 795–804 (1999).
63. J. Vargas, J. A. Quiroga, C. O. S. Sorzano, J. C. Estrada, and J. M. Carazo, "Two-step interferometry by a regularized optical flow algorithm," Opt. Lett. **36**(17), 3485–3487 (2011).
64. K. Fukushima, "Neocognitron: A Self-organizing Neural Network Model for a Mechanism of Pattern Recognition Unaffected by Shift in Position," Biol. Cybernetics **36**, 193-202 (1980).

65. S. Cho, "A Neural Network for Denoising Fringe Patterns with Nonuniformly Illuminating Background Noise," J. Korean Phys. Soc. **75**(6), 454-459 (2019).
66. B. Lin, S. Fu, C. Zhang, F. Wang and Y. Li, "Optical fringe patterns filtering based on multi-stage convolution neural network," Opt. Lasers Eng. **126**, 105853 (2020).
67. K. Yan, J. Shi, T. Sun, J. Li and Y. Yu, "Fringe pattern filtering using convolutional neural network," Proc. SPIE **11205**, 112050O (2019).
68. K. Yan, Y. Yu, C. Huang, L. Sui, K. Qian and A. Asundi, "Fringe pattern denoising based on deep learning," Opt. Commun. **437**, 148-152 (2019).
69. W. Xiao, Q. Wang, F. Pan, R. Cao, X. Wu, and L. Sun, "Adaptive frequency filtering based on convolutional neural networks in off-axis digital holographic microscopy," Biomed. Opt. Express **10**(4), 1613-1626 (2019).
70. X. He, C. V. Nguyen, M. Pratap, Y. Zheng, Y. Wang, D. R. Nisbet, R. J. Williams, M. Rug, A. G. Maier, and W. M. Lee, "Automated Fourier space region-recognition filtering for off-axis digital holographic microscopy," Biomed. Opt. Express **7**(8), 3111-3123 (2016).
71. P. Memmolo, V. Renò, E. Stella, and P. Ferraro, "Adaptive and automatic diffraction order filtering by singular value decomposition in off-axis digital holographic microscopy," Appl. Opt. **58**(34), G155-G161 (2019).
72. S. Feng, Q. Chen, G. Gu, T. Tao, L. Zhang, Y. Hu, W. Yin and C. Zuo, "Fringe pattern analysis using deep learning," Adv. Photon. **1**(2), 025001 (2019).
73. J. Shi, X. Zhu, H. Wang, L. Song, and Q. Guo, "Label enhanced and patch based deep learning for phase retrieval from single frame fringe pattern in fringe projection 3D measurement," Opt. Express **27**(20), 28929-28943 (2019).
74. S. Van der Jeught and J. J. J. Dirckx, "Deep neural networks for single shot structured light profilometry," Opt. Express **27**(12), 17091-17101 (2019).
75. H. Yu, X. Chen, Z. Zhang, C. Zuo, Y. Zhang, D. Zheng, and J. Han, "Dynamic 3-D measurement based on fringe-to-fringe transformation using deep learning," Opt. Express **28**(7), 9405-9418 (2020).
76. H. Nguyen, N. Dunne, H. Li, Y. Wang, and Z. Wang, "Real-time 3D shape measurement using 3LCD projection and deep machine learning," Appl. Opt. **58**(26), 7100-7109 (2019).
77. K. Wang, Y. Li, Q. Kemao, J. Di, and J. Zhao, "One-step robust deep learning phase unwrapping," Opt. Express **27**(10), 15100-15115 (2019).
78. G. E. Spoorthi, S. Gorthi, R. K. Sai and S. Gorthi, "PhaseNet: A Deep Convolutional Neural Network for Two-Dimensional Phase Unwrapping," IEEE Signal Processing Letters **26**(1), 54-58 (2019).
79. T. Zhang, S. Jiang, Z. Zhao, K. Dixit, X. Zhou, J. Hou, Y. Zhang, and C. Yan, "Rapid and robust two-dimensional phase unwrapping via deep learning," Opt. Express **27**(16), 23173-23185 (2019).
80. J. Zhang, X. Tian, J. Shao, H. Luo, and R. Liang, "Phase unwrapping in optical metrology via denoised and convolutional segmentation networks," Opt. Express **27**(10), 14903-14912 (2019).
81. G. Dardikman-Yoffe, D. Roitshtain, S. K. Mirsky, N. A. Turko, M. Habaza, and N. T. Shaked, "PhUn-Net: ready-to-use neural network for unwrapping quantitative phase images of biological cells," Biomed. Opt. Express **11**(2), 1107-1121 (2020).
82. Y. Qin, S. Wan, Y. Wan, J. Weng, W. Liu, and Q. Gong, "Direct and accurate phase unwrapping with deep neural network," Appl. Opt. **59**(24), 7258-7267 (2020).
83. M. Cywińska, F. Brzeski, W. Krajnik, K. Patorski, C. Zuo, and M. Trusiak, „DeepDensity: Convolutional neural network based estimation of local fringe pattern density," Opt. Lasers Eng. **145**, 106675 (2021).
84. L. Tian, C. Tang, M. Xu, and Z. Lei, "Accurate and efficient extraction of fringe orientation from the poor-quality ESPI fringe pattern with a convolutional neural network," Appl. Opt. **58**(27), 7523-7530 (2019).
85. M. Trusiak, Ł. Służewski, and K. Patorski, "Single shot fringe pattern phase demodulation using Hilbert-Huang transform aided by the principal component analysis," Opt. Express **24**(4), 4221-4238 (2016)
86. K. Dabov, A. Foi, V. Katkovnik and K. Egiazarian, "Image denoising by sparse 3D transform-domain collaborative filtering," IEEE Trans. Image Process. **16**(8), 2080-2095 (2007).
87. M. Cywińska, M. Trusiak, and K. Patorski, "Automatized fringe pattern preprocessing using unsupervised variational image decomposition," Opt. Express **27**(16), 22542-22562 (2019).
88. P. Gocłowski, M. Trusiak, A. Ahmad, A. Styk, V. Mico, B. S. Ahluwalia, and K. Patorski, „Automatic fringe pattern enhancement using truly adaptive period-guided bidimensional empirical mode decomposition," Opt. Express **28**(5), 6277-6293 (2020).
89. M. Rogalski, M. Pielach, A. Cicone, P. Zdańkowski, L. Stanaszek, K. Drela, K. Patorski, B. Lukomska, and M. Trusiak, "Tailoring 2D fast iterative filtering algorithm for low-contrast optical fringe pattern preprocessing," Opt. Lasers Eng. **155**, 107069 (2022).
90. V. Dubey, D. Popova, A. Ahmad, G. Acharya, P. Basnet, D.S. Mehta, and B.S. Ahluwalia, "Partially spatially coherent digital holographic microscopy and machine learning for quantitative analysis of human spermatozoa under oxidative stress condition," Sci. Rep. **9**(1), 3564 (2019)
91. M. Cywińska, M. Rogalski, F. Brzeski, K. Patorski, M. Trusiak, „DeepOrientationNetModel" https://github.com/MariaSi1/DeepOrientationNetModel